\documentstyle[prl,preprint,aps,epsf]{revtex}
\begin{document}
\preprint{ ETH-TH/96-17}
\draft
\title{Peaks above the Harrison-Zel'dovich spectrum due to the Quark-Gluon to
Hadron Transition}
\author{Christoph Schmid, Dominik J. Schwarz, and Peter Widerin\thanks{
e-mail: chschmid, dschwarz, widerin, all @itp.phys.ethz.ch}}
\address{Institut f\"ur 
Theoretische Physik, ETH-H\"onggerberg, CH-8093 Z\"urich}
\date{\today}
\maketitle
\tighten
\begin{abstract}
The quark-gluon to hadron transition affects the evolution of cosmological 
perturbations. If the phase transition is first order, the sound speed
vanishes during the transition, and density perturbations fall freely.
This distorts the primordial Harrison-Zel'dovich spectrum of density 
fluctuations below the Hubble scale at the transition. Peaks are produced,
which grow at most linearly in wave\-number, both for the hadron-photon-lepton
fluid and for cold dark matter. For cold dark matter which is kinetically
decoupled well before the QCD transition clumps of masses below $10^{-10} 
M_\odot$ are produced.
\end{abstract}
\narrowtext

QCD makes a transition from a quark-gluon plasma
at high temperatures to a hadron gas at low temperatures. 
Lattice QCD simulations give a transition temperature 
$T_\star \sim 150$ MeV and indicate a 
first-order phase transition for the physical values of the u,d,s-quark masses
\cite{lattice}. The relevance of the QCD transition for cosmology,
especially for big-bang nucleosynthesis \cite{BBN}, has been discussed
before, but the focus was on effects of bubble formation 
\cite{bubbles,Christiansen}.
In this paper we look at matter averaged over scales 
$\lambda$ much larger than the bubble separation. We
show that for a first-order phase transition the sound speed $c_s
= (\partial p/\partial \rho)_s^{1/2}$ drops to zero for
these wavelengths
when the transition temperature $T_\star$ is reached, 
stays zero for the entire 
time until the phase transition is completed, and
afterwards suddenly rises back to $c_s \approx
c/\sqrt{3}$. In contrast the pressure
stays positive and varies continuously, although it goes below the radiation
fluid value $p = \rho/3$. Since $c_s$ is zero
during the transition, there are no pressure perturbations, no pressure 
gradients, no restoring forces.
Pre-existing cosmological perturbations, generated by
inflation \cite{infl} with a Harrison-Zel'dovich spectrum \cite{HZ},
go into free fall. 
The superhorizon modes (at the time of the transition) remain 
unaffected. The subhorizon modes
develop peaks in $\delta\!\rho/\rho$ which grow with wavenumber $k 
> k_\star$, where $k_\star^{\rm phys}\sim$ Hubble rate $H$ at the end
of the QCD transition. The details of this growth depend on the QCD equation of 
state near $T_\star$. We analyze two cases: 
First we use the bag model \cite{DeGrand}, which gives a simple 
parameterization and allows a
simple discussion of the effects. It gives a maximal latent heat, and
produces peaks in $\delta\!\rho/\rho$ which grow linearly in $k$.
Next we use lattice QCD results \cite{lattice,Karsch,Beinlich}, 
which indicate a smaller latent heat, 
and we fit $s/T^3 = C_1 + C_2 (1 - T_\star/T)^{1/3}$ above $T_\star$. This 
produces peaks in $\delta\!\rho/\rho$ which grow as $k^{3/4}$.
 
The sound speed (for wavelength $\lambda$ much larger than the 
bubble separation), $c_s = (\partial p/\partial \rho)_s^{1/2}$, must be zero 
during a first-order phase transition of a fluid with negligible chemical 
potential, since the fluid must obey
\begin{equation}
\label{2nd}
\rho + p = T {dp\over dT} \ ,
\end{equation}
according to the second law of thermodynamics. Because the energy
density $\rho$ is discontinuous in temperature at $T_\star$ for a first-order
phase transition, the pressure $p$ must be continuous with
a discontinuous slope. As the universe expands at fixed temperature $T_\star$
during the phase transition, $\rho$ as a function of time slowly decreases
from $\rho_+(T_\star)$ to $\rho_-(T_\star)$, $p$ stays constant at 
$p(T_\star)$, and therefore $c_s$ is zero during the whole time of the
phase transition.

The interaction rates in the QCD-photon-lepton fluid
are much larger than the Hubble rate, $\Gamma/H\gg 1$,
therefore we are very close to thermal and chemical equilibrium, 
the QCD transition
is very close to a reversible thermodynamic transformation, and 
the entropy in a comoving volume is approximately conserved.  
Estimates show that supercooling, hence entropy production,
is negligible, $(T_\star - T_{\rm supercooling})/T_\star
\sim 10^{-3}$ \cite{supercool}.
Bubble formation during the QCD phase transition is unimportant for our
analysis, estimates give a bubble separation $\ell_b \sim 1$ cm
\cite{Christiansen}, while the Hubble radius at the QCD transition
is $R_H \sim 10$ km, therefore $\ell_b/R_H \sim 10^{-6}$.
We shall analyze perturbations with $\lambda \gg \ell_b$.

In the bag model \cite{DeGrand} it is assumed that for 
$T>T_\star$ the quark-gluon plasma (QGP) obeys
\begin{equation}
\label{p}
p_{\rm QGP}(T) = p_{\rm QGP}^{\rm ideal}(T) - B \ ,
\end{equation}
where $p_{\rm QGP}^{\rm ideal}(T) = (\pi^2/90) g_{\rm QGP}^* T^4$, $g^*$ is the
effective number of relativistic helicity states, and $B$
is the bag constant. 
We include u,d-quarks and gluons in the quark-gluon plasma, $\gamma,e,\mu$,
and 3 neutrinos in the photon-lepton fluid ($\gamma{\rm L}$), and  
for $T<T_\star$ we have a hadron gas (HG) of pions. 
We treat the pions as massless and ideal, because their contribution
is small anyway, $g_{\rm HG}^*/g^*_{\rm QGP} = 3/37$ and 
$g_{\rm HG}^*/g^*_{\gamma\rm L} = 3/14.25$. 
$\rho$ follows from 
Eq.~(\ref{p}) via the second law of thermodynamics, Eq.~(\ref{2nd}), and 
$s$ from $s = dp/dT$. This gives for the quark-gluon plasma
\begin{eqnarray}
\label{r}
\rho_{\rm QGP}(T) &=& \rho_{\rm QGP}^{\rm ideal}(T) + B \\
s_{\rm QGP}(T) &=& s_{\rm QGP}^{\rm ideal}(T) \ .
\end{eqnarray}
The bag constant is determined by the critical temperature $T_\star$ via
$p_{\rm QGP}(T_\star) = p_{\rm HG}(T_\star)$.

The latent heat, $L \equiv T_\star \Delta s$, should be compared with
the difference in entropy between an ideal HG and an
ideal QGP. This defines the ratio 
$R_L \equiv L/(T_\star \Delta s)^{\rm ideal}$. The bag model gives $R_L = 1$.
Lattice QCD indicates that the transition is first order both for
quenched QCD (no dynamical quarks) \cite{Karsch} and for QCD with three
quarks and physical masses \cite{lattice}. For the latter case neither the
value of the latent heat nor the equation of state are available. 
Quenched QCD gives $L/T_\star^4 \approx 1.4$ \cite{Beinlich}, which implies 
$R_L \approx 0.2$. We fit the shape of the QCD entropy to quenched 
QCD data \cite{Karsch} for $T >  T_\star$ by
\begin{equation}
\label{fit}
{s^{\rm fit}_{\rm QGP}\over s^{\rm ideal}_{\rm QGP}} = 
1 + \left[\left(1 - {T_\star\over T}\right)^\gamma -1\right]
{\Delta g^*\over g_{\rm QGP}^*}\left(1 - R_L\right) \ ,
\end{equation}
where $\Delta g^* \equiv g^*_{\rm QGP} - g^*_{\rm HG}$ and $R_L = 0.2$. 
A good fit for our 
purpose is obtained for $\gamma \in (0.3, 0.4)$. We fix $\gamma = 1/3$.
 
\begin{figure}
\begin{center}
\epsfig{figure=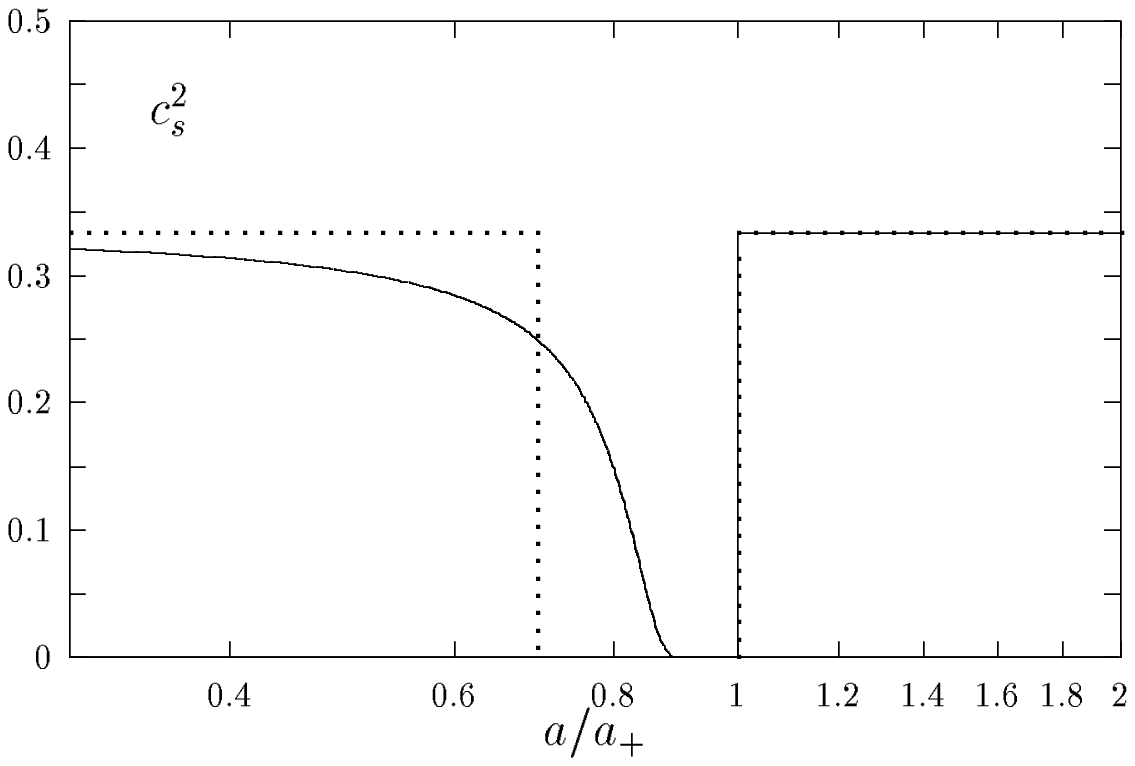,width=0.7\linewidth}
\end{center}
\vspace{-7pt}
\caption{
The sound speed $c_s^2 = (\partial p/\partial\rho)_s$ during the
QCD transition for the bag model (dotted line)
and for the lattice QCD fit (full line).}
\label{fig1}
\end{figure}

The growth of the scale factor during the $c_s^2 = 0$ part of the QCD
transition, $a_+/a_-$, follows from the conservation of entropy in a comoving 
volume, 
\begin{equation}
{a_+\over a_-} = \left[1 + R_L {\Delta g^*\over g_{\rm after}^*}
\right]^\frac13 \approx
\left\{ \begin{array}{ll}
               1.4 & \qquad R_L = 1 \\ 
               1.1 & \qquad R_L = 0.2
        \end{array}\right. ,
\end{equation}
taking into account photons, leptons, and hadrons in $g^*_{\rm after}$.
Fig.~\ref{fig1} shows the evolution of the sound 
speed with the scale factor $a$.
Above $T_\star$, the sound speed in the bag model
has the value for an ultrarelativistic ideal gas, $c_s = 1/\sqrt{3}$, 
because the bag constant drops out when forming $dp/dT$ and $d\rho/dT$ 
in Eqs.~(\ref{p}) and (\ref{r}).
The sound speed vanishes for about a third of a Hubble time for $R_L = 1$
and for a tenth of a Hubble time if $R_L = 0.2$. 
The pressure does not drop all the
way to zero, it drops to $p_{\gamma \rm L}(T_\star) + p_{\rm
HG}(T_\star)$.

The evolution of linear cosmological perturbations through 
the QCD transition is
analyzed in the longitudinal sector
(density perturbations) for perfect fluids. We choose a slicing 
$\Sigma$ of space-time with unperturbed mean extrinsic curvature,
$\delta[\mbox{tr} K_{ij}(\Sigma)] = 0$. This implies that our fundamental
observers, which are defined to be at rest on the slice $\Sigma$, 
$\underline{u}(\mbox{obs}) = \underline{n}(\Sigma)$, have relative velocities,
which in the mean over all directions follow an unperturbed Hubble
flow.
If the coordinate choice (gauge choice) is such that the time coordinate $t$
is constant on the slices $\Sigma$, the gauge is fixed to be the 
uniform expansion (Hubble) gauge \cite{Bardeen}.
As fundamental evolution equations for each fluid we have $\nabla_{\mu}
T^{\mu\nu} = 0$, i.e. 
the continuity equation and (in the longitudinal sector)
the 3-divergence of the Euler equation
of general relativity,
\begin{eqnarray}
\label{C}
\partial_t \epsilon &=& - 3 H(\epsilon + \pi) - \bigtriangleup \psi -
3 H (\rho + p)\alpha \\
\label{E}
\partial_t \psi &=& - 3 H \psi - \pi - (\rho + p)\alpha \ ,
\end{eqnarray}
where $\epsilon \equiv \delta\!\rho$, $\pi \equiv \delta\! p$,
$\rho \equiv \rho_0$, $p \equiv p_0$, 
$\vec{\nabla} \psi \equiv \vec{S} =$ momentum density (Poynting vector),
$\alpha =$ lapse function. The system of dynamical equations is
closed by Einstein's $R_{\hat{0}\hat{0}}$-equation, the 
general relativistic version of Poisson's equation,
\begin{equation}
\label{R}
(\bigtriangleup + 3 \dot{H})\alpha = 4\pi G (\epsilon + 3 \pi) \ ,
\end{equation}
together with the equation of state.
Equations (\ref{C}) -- (\ref{R}) define our general relativistic Cauchy
problem in linear perturbation theory in the longitudinal sector with initial
data $(\epsilon, \psi)$ freely chosen on $\Sigma_i$. These three equations are
the Jeans equations extended to general relativity in the 
longitudinal sector. In all
three of them the mean over all directions is taken. This fact matches our
slicing condition that in the mean over all directions the relative velocity
of our fundamental observers is unperturbed. Therefore the uniform expansion
(Hubble) gauge could be called the 'longitudinal Jeans gauge'. 

It is convenient to work with the dimensionless variables $\delta \equiv
\epsilon/\rho$ (density contrast), $\hat{\psi} \equiv k^{\rm phys} \psi/\rho$
($\sim$ peculiar velocity) and with conformal time, $( )^\prime
\equiv \partial_{\eta} \equiv a \partial_t$. In our numerical analysis
we have used the exact general relativistic equations, but it is instructive to
look at the subhorizon approximation, $\lambda^{\rm phys} \ll H^{-1}$,
where one can drop $\dot{H}$ in the general relativistic Poisson equation
(\ref{R}) and the time dilation term (last term)
in the continuity equation (\ref{C}).
Furthermore, if we take the limit in which the QCD transition
time is much shorter than the Hubble time, $(t_+ - t_-) \ll H^{-1}$, 
and if we integrate the equations during
a correspondingly short time interval, we can drop the remaining terms
proportional to $H$, and Eqs. (\ref{C}) -- (\ref{R}) simplify to
\begin{eqnarray}
\delta^\prime &=& k \hat{\psi} \nonumber \\
\label{eqns}
\hat{\psi}^\prime &=& - c_s^2 k \delta - (1+w)k\alpha \\
\left(k\over aH\right)^2 \alpha &=& -\frac32 (1+3 c_s^2)\delta \nonumber \ ,
\end{eqnarray}
where $w \equiv p/\rho$.
With these approximations the evolution of a mode $k$ of cosmological
perturbations can be solved analytically.

The origin of large peaks in $\delta\!\rho/\rho$
for $k \gg k_\star$, where $k_\star^{\rm phys} \sim  H$ at the transition, is 
easily understood in the bag model. 
For the dynamics of the radiation fluid (QCD, photons, leptons)
one can neglect cold dark matter
(CDM) since 
$\rho_{\rm CDM}/\rho_{\rm RAD} = a/a_{\rm equality} \approx 10^{-8}$.
The radiation fluid in each mode makes standing acoustic oscillations 
before and after
the QCD transition with gravity negligible and with equal
amplitudes of $\delta$ and $\sqrt{3}\hat{\psi}$.
The solution before the QCD transition is $\delta(\eta) =
A_{\rm in} \cos[\omega(\eta - \eta_-) - \varphi_-]$ and $\hat{\psi} =
\delta^\prime/k$, where $\omega = k c_s = k/\sqrt{3}$, and
$\varphi_-$ is the phase of the acoustic oscillation at $\eta_-$, i.e.
at the beginning of the QCD transition.
During the QCD transition the sound speed is zero, there are
no restoring forces from pressure gradients, the radiation fluid  
goes into free fall. But during this free fall gravity is again 
negligible for the radiation fluid, if $(t_+ - t_-) \ll H^{-1}$.
This is inertial motion in the sense of Newton. 
The peculiar velocity is 
constant in time, $\hat{\psi}(\eta) = \hat{\psi}_-$, and the density
contrast grows linearly in time with a slope $k$, $\delta(\eta) =
\delta_- + k(\eta - \eta_-) \hat{\psi}_-$.
Thus, the final amplitude $A_+$ grows linearly
in $k$ modulated by $\sin(\varphi_-)$, which produces peaks
in the spectrum. The height of these peaks is 
\begin{equation}
\label{k1}
\left. A_+\over A_{\rm in}\right|_{\rm peaks} \!\!\!= 
{(\eta_+ - \eta_-)k\over\sqrt{3}}
\equiv {k \over k_1}\ ,
\end{equation}
for $k \gg k_1$.
The usual free fall growing mode at subhorizon scales behaves
totally different: 
$\hat{\psi}/\delta \sim H/k^{\rm phys} \ll 1$ and $\delta_+/\delta_- =
(\eta_+/\eta_-)^2 \approx 2$. In our case the initial peculiar velocity
$\hat{\psi}$ from the acoustic oscillations of the radiation fluid 
is enormously larger than in the usual free fall growing mode 
for $k \gg k_\star$. 
COBE observations \cite{COBE}
normalize the subhorizon spectrum of density perturbations 
as $(\delta\!\rho/\rho)_{\lambda} 
\sim 10^{-4}$, if there is no tilt in the spectrum.
Hence in the bag model 
radiation-fluid modes with $k/k_1 \gtrsim 10^4$ go nonlinear by
the end of the QCD transition.

\begin{figure}
\begin{center}
\epsfig{figure=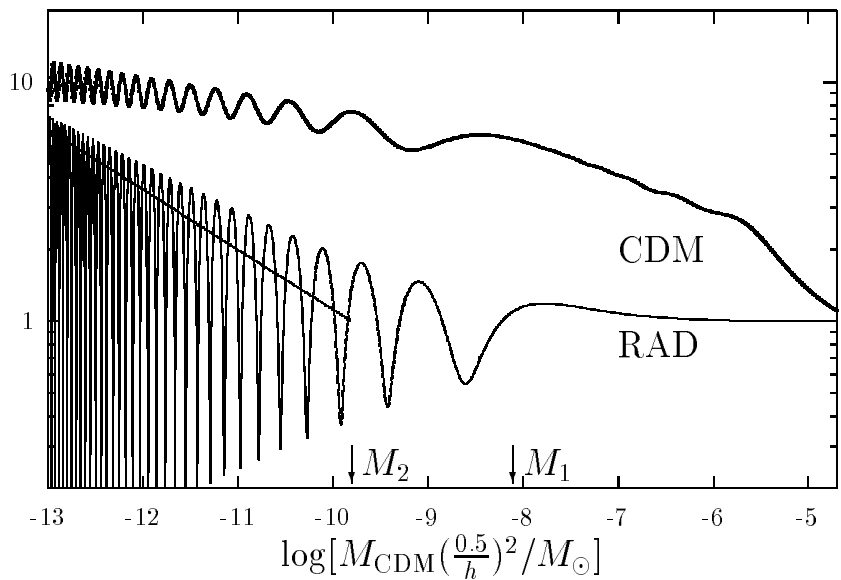,width=0.7\linewidth}
\end{center}
\vspace{-7pt}
\caption{
The modifications of the CDM density contrast $A_{\rm CDM} \equiv |
\delta_{\rm CDM}| (T_\star /10)$ and
of the radiation fluid amplitude $A_{\rm RAD} \equiv (\delta_{\rm RAD}^2 + 3
\hat{\psi}_{\rm RAD}^2)^{1/2}$  due to the QCD transition (lattice
QCD fit). 
Both quantities are 
normalized to the pure Harrison-Zel'dovich radiation amplitude. On the 
horizontal axis the wavenumber $k$ is represented by the CDM mass contained
in a sphere of radius $\pi/k$.} 
\label{fig2}
\end{figure}

\begin{figure}
\begin{center}
\epsfig{figure=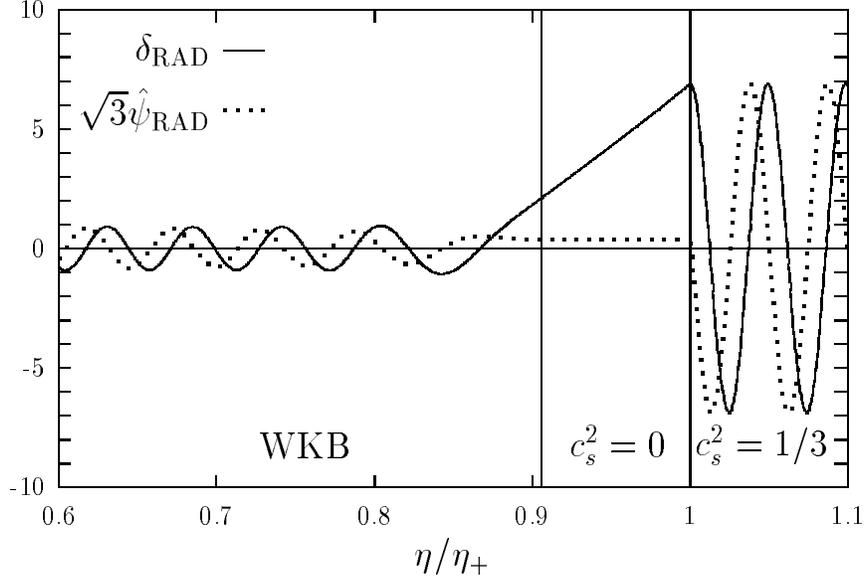,width=0.7\linewidth}
\end{center}
\vspace{-7pt}
\caption{
The evolution in conformal time $\eta$ of the density contrast ($\delta_{\rm
RAD}$) and the peculiar velocity ($\sim \hat{\psi}_{\rm RAD}$)
of the radiation fluid for the highest peak of Fig.~2 
in uniform expansion (Hubble) gauge. During the
QCD transition in the lattice QCD fit 
--- marked by the 2 vertical lines --- the velocity stays
approximately constant and the density contrast grows linearly.
The amplitude in the WKB regime is normalized to $1$ long before 
the transition.} 
\label{fig3}
\end{figure}

Our numerical results 
for the spectrum of density perturbations from the lattice QCD
fit Eq.~(\ref{fit}) are given in Fig.~\ref{fig2}. 
We show the enhancement of the amplitude $A_{\rm RAD} 
\equiv (\delta_{\rm RAD}^2 + 3 \hat{\psi}_{\rm RAD}^2)^{1/2}$ of 
the acoustic oscillations of the
radiation fluid after the transition compared 
to the amplitude without transition. For CDM 
we show the amplitude $A_{\rm CDM} \equiv |\delta_{\rm CDM}|$ 
at $T_\star/10$ compared to $A_{\rm RAD}$ without transition. 
In both cases we obtain
peaks over the Harrison-Zel'dovich spectrum of primordial adiabatic
density fluctuations. 
The modes $k$ (horizontal axis) are labeled
by the CDM mass contained in a sphere of radius $\lambda/2 = \pi/k$.
The positions of the first few peaks and dips are the same for the
lattice QCD fit and for the bag model. In both cases the beginning of the 
peak-dip structure is at $k_1$, see Eq.~(\ref{k1}), which is $\sim k_\star$. 
$k_1$ corresponds to
$M_1^{\rm CDM}  \approx 9\times 10^{-9} M_\odot$.
The peaks grow as $(k/k_2)^{3/4}$ for $k \gg k_2$, see Eq.~(\ref{13}).
The radiation energy inside $\lambda_1/2$ is $\sim 1 M_\odot$,
but it gets redshifted as 
$M_{\rm RAD}(a) \sim (a_{\rm equality}/a) M_{\rm CDM}$.
The time evolution of a subhorizon mode is shown in Fig.~\ref{fig3}.
During $c_s=0$ $\delta_{\rm RAD}$ grows linearly as in the bag model. 
Above $T_\star$
the evolution differs because $c_s < 1/\sqrt{3}$. 

The slower rise in $k$ for the lattice QCD fit can be understood by a 
WKB analysis. With the same approximations as above, Eqs.~(\ref{eqns})
reduce to 
\begin{equation}
\label{12}
\delta'' + c_s^2 k^2 \delta = 0 \ . 
\end{equation}
Under the WKB condition, $|dc_s/d\eta|/c_s \ll \omega = c_s k$, the solution
reads $\delta = A_{\rm in} (3 c_s^2)^{-1/4} \cos(k \int c_s d\eta)$.
Note that the peculiar velocity
$(\sim \hat{\psi})$ decreases with $c_s^{1/2}$, which can be seen in 
Fig.~\ref{fig3}.
Just above $T_\star$ the sound speed may be approximated by 
$c_s \propto (\eta_- - \eta)$. 
This gives the solutions of Eq.~(\ref{12}) as $\delta \propto z^{1/4} 
J_{\pm 1/4}(z)$ with $z \equiv c_s k (\eta_- - \eta)/2$.
The normalization for large $k$ is provided by the WKB solution.
At $\eta_-$ one matches the linearly growing solution in the regime of 
vanishing sound speed ($\eta_- < \eta < \eta_+$). The final
amplification is 
\begin{equation}
\label{13}
\hspace{-1pt} \left. A_+\over A_{\rm in}\right|_{\rm peaks} \!\!\! = 
{\sqrt{\pi}\over 2\Gamma (\frac54 )}
\left|{c_s'(\eta_-)\over 3k}\right|^{\frac14}\!\!
(\eta_+ - \eta_-)k \equiv \left(k\over k_2\right)^{\frac34} , 
\end{equation}
for $k \gg k_2$.
$k_2$ depends on the duration of the $c_s = 0$ regime and thus depends 
on $R_L$.  For $R_L = 0.2$ it corresponds to a mass $M_2 = 2 \times 
10^{-10} M_\odot$, which is indicated in 
Fig.~\ref{fig2} together with the asymptotic envelope $(k/k_2)^{3/4}$.
Without tilt COBE normalized modes with
$k/k_2 \gtrsim 10^5$ go nonlinear by the end
of the QCD transition.
The perturbations in the radiation fluid will get wiped out 
by collisional damping from neutrinos 
at temperatures below $T_\star$ but above $1$ MeV. 

For cold dark matter we consider any non-relativistic 
matter which decouples kinetically well before the QCD transition. 
The neutralino, most likely the lightest
supersymmetric particle \cite{Griest}, is weakly interacting. 
Thus it decouples kinetically
around $T \sim 1$ MeV and is excluded to make our CDM. 
It would belong to the radiation fluid at $T_\star$.
Candidates for our CDM are axions
or primordial black holes.

CDM falls into the gravity wells generated during the transition
by the radiation fluid. In the bag model this leads to peaks in CDM which
grow linearly for $k \gg k_1$, i.e. $\delta^{\rm CDM}_+ - 
\delta^{\rm CDM}_- = [H(t_+ - t_-)/2]^2 \delta^{\rm RAD}_+$. After the
transition $\delta^{\rm CDM}$ grows logarithmically. 
For our lattice QCD fit the CDM peaks are shown 
in Fig.~\ref{fig2}. An analytic analysis of the CDM evolution will be
presented in a longer paper.

The implications of these peaks above the Harrison-Zel'dovich spectrum 
generated in a standard scenario with a first order QCD transition are:\\
{\bf 1)} For CDM which is kinetically decoupled well before
the QCD transition clumps with $M_{\rm CDM} \lesssim 10^{-10} M_\odot$
are produced. They go nonlinear after equality and virialize 
by violent gravitational relaxation. Assuming a COBE normalized spectrum 
with tilt $n -1 = 0 (0.2)$ and $3\sigma$ peaks the size of 
$10^{-10} M_\odot$ clumps is $\approx 14$ AU ($ 1$ AU).\\
{\bf 2)} Big-Bang Nucleosynthesis will not be affected by
nonlinear acoustic oscillations of the radiation fluid for 
$k/k_1 \gtrsim 10^4$ (bag model) resp.\ $10^6$ (lattice QCD fit), 
because they are wiped out by collisional damping
from neutrinos before Big-Bang Nucleosynthesis.\\
{\bf 3)} Primordial black hole formation is unlikely, 
because the nonlinear acoustic oscillations in the radiation fluid 
are generated far below the Hubble scale at the QCD transition.
This is in contrast to Ref.~\cite{J}.

We like to thank J. A. Bardeen, P. Jetzer, F. Karsch, V. Mukhanov,
J. Silk, and N. Straumann for helpful
discussions.
C. S. thanks the Center for Particle Astrophysics in Berkeley and the
Fermilab Astrophysics Center for hospitality.
D. S. and P. W. thank the Swiss National Science Foundation for financial
support.

\end{document}